\documentstyle[12pt]{article}

\parskip=\medskipamount
\begin{document}
\baselineskip=20pt

\begin{center}
{ \large \bf Isospin correlations in high-energy heavy-ion collisions
  }\\
\vspace{2.cm}
{ \normalsize \bf M. Martinis, \footnote { e-mail address:
martinis@thphys.irb.h
r}
 V. Mikuta-Martinis, \footnote {e-mail address: mikutama@thphys.irb.hr}
 A. \v Svarc,
and J. \v Crnugelj \footnote
{e-mail address:crnugelj@thphys.irb.hr }} \\
\vspace{0.5cm}
Department of Physics, \\
Theory Division, \\
Rudjer Bo\v skovi\' c Institute, P.O.B. 1016, \\
41001 Zagreb,CROATIA \\
\end{center}
{ \large \bf Abstract}
\vspace{0.5cm}
\baselineskip=20pt

We study the posibility of large isospin fluctuations
in high-energy heavy-ion collisions
by assuming that pions are produced semiclassically both directly
and in pairs through the isovector channel.

The leading-particle effect and the factorization property of the
scattering amplitude in the impact parameter space are used to define
the classical pion field. In terms of the joint probability function
$P_{II_{3}}(n_{0},n_{ \_})$ for producing $n_{0}$ neutral and
$n_{ \_}$ negative pions from a definite isospin state $II_{3}$
of the incoming leading-particle system we calculate the two pion
correlation parameters $f_{2,n_{ \_}}^{0}$ and  the
average number of neutral pions $( \langle n_{0} \rangle_{n_{ \_}})$ as
a function of negative pions $(n_{ \_})$ produced.
We show that only direct
production of pions without isovector pairs leads to large isospin
fluctuations.

\vspace{1cm}

\newpage
\baselineskip=24pt
\section{Introduction}
In recent years several cosmic-ray experiments [1] have reported evidence
for the existence of Centauro events characterized by an anomalously large
number of charged pions in comparison with the number of neutral pions,
indicating that there should exist a  strong long-range correlation
between two types of the pions. The negative results of the accelerator
searches for the Centauro events at CERN (2,3,4) have suggested that
threshold for their production must be larger then 900 GeV.

 Such long-range correlations are possible
if pions are produced semiclassicaly and constrained by global
 conservation of isospin [5--10].

Although the actual dynamical mechanism of the
production of a classical pion field in the course
of a high-energy collision is not known, there exist a
number of interesting theoretical speculations [11--16] that localized
regions of misaligned chiral vacuum might occur in
ultrahigh-energy hadronic
and heavy-ion collisions. These regions become coherent sources of a
classical pion field. In early models [5,6], however, the coherent
production of pions was taken for granted and
considered as a dominant mechanism.

These models  also predict strong negative correlations between the number
of neutral and charged pions. In fact, the exact conservation of isospin
in a pion uncorrelated jet model is known [7,8] to
give the same pattern of neutral/charged fluctuations
as observed in Centauro events. This strong negative
neutral-charged correlation is believed to be a
general property of the direct pion emission in
which the cluster formation ( or the short-range correlation between pions )
is not taken into account [9,10,17,18,19].

In this paper, following the approach of our earlier paper (18)
 we consider the leading-particle effect as a  source
of a classical pion field in the impact parameter space. Pions are
assumed to be produced from a definite isospin state of the
incoming two-particle system both directly and in
isovector pairs.

Coherently emitted isovector clusters decay subsequently into
pions outside the region of $ \pi \pi$ interaction. We discuss the
behavior of the joint probability distribution
$P_{II_{3}}(n_{0},n_{ \_})$ of
neutral and charged pions, the variation of the
two-pion correlation functions $f_{2,n_{ \_}}^{0}$
and the average number of neutral pions
$( \langle n_{0} \rangle_{n_{ \_}} )$ as a function of
the number of negative pions $(n_{ \_})$ produced.

We support the conclusion of (18,19) that the large isospin
fluctuations are consequence of a direct production of pions.

\newpage
\baselineskip=24pt
\section{The Eikonal $S$ matrix with Isospin}
At high energies most of the pions are
produced in the central region. To isolate the central production,
we adopt high-energy longitudinally dominated kinematics, with
two leading particles retaining a large fraction of their incident
momenta. We assume that the collision energy is large enough
so that the central region is free of baryons.

The basic assumption of the independent pion-emission model,
neglecting the isospin for a moment, is the factorization of the
scattering amplitude $ T_{n}( s, \vec{b}; 1 \ldots n )$ in the
$b$ space:
\begin{equation}
T_{n}( s, \vec{b}; 1 \ldots n) = 2sf(s, \vec{b}) \frac{ \textstyle
i^{n-1}}{ \textstyle \sqrt{n!}} \prod_{i=1}^{n} J( s, \vec{b}; q_{i}),
\end{equation}
where, owing to unitarity,
\begin{equation}
\mid f(s, \vec{b}) \mid^{2} = e^{ \textstyle - \overline{n}(s, \vec{b})}
\end{equation}
and
\begin{equation}
\overline{n}(s, \vec{b}) = \int dq \mid J(s, \vec{b};q) \mid^{2}
\end{equation}
denotes the average number of emitted pions at a given impact
parameter $b$.
The function $ \mid J(s, \vec{b}; q) \mid^{2}$, after the integration over b,
controls the shape of the single-particle inclusive distribution.
A suitable choice of this function also
guarantees that the energy and the momentum
are conserved on the average during the collision.

The inclusion of isospin in this model is straightforward [6].
The factorization of $T_{n}$ in the form (1) is a consequence of
the pion field satisfying the equation of motion
\begin{equation}
( \Box + \mu^{2}) \vec{ \pi}( s,\vec{b}; x) = \vec{j}(s,
\vec{b};x),
\end{equation}
where $ \vec{j}$ is a classical source related to $ \vec{J}(s,
\vec{b};q )$ via the Fourier transform
\begin{equation}
\vec{J}(s, \vec{b};q) = \int d^{4}x e^{ iqx}
\vec{j}(s, \vec{b};x).
\end{equation}

The standard solution of Eq.(4) is usually given in terms of in- and
out-fields that are connected by the unitary
$S$ matrix $ \hat{S}( \vec{b},s)$ as follows:
\begin{equation}
\vec{ \pi}_{out} = \hat{S}^{ \dagger} \vec{ \pi}_{in}
\hat{S} = \vec{ \pi}_{in} + \vec{ \pi}_{classical},
\end{equation}
where
\begin{equation}
\vec{ \pi}_{classical} = \int d^{4}x' \Delta(x-x'; \mu) \vec{j}(s,
\vec{b};x').
\end{equation}

The $S$ matrix following from such a classical source
is still an operator in the space of pions.
Inclusion of isospin requires $ \hat{S}(s, \vec{b})$
to be also a matrix in the isospace of the leading particles.

The coherent production of isovector clusters of pions
is described by the following $S$ matrix:
\begin{equation}
\hat{S}(s, \vec{b}) = \int d^{2} \vec{e} \mid \vec{e}
\rangle D( \vec{J}; s, \vec{b}) \langle \vec{e} \mid,
\end{equation}
where $ \mid \vec{e} \, \rangle $ represents the isospin-state
vector of the two-leading-particle system.
The quantity $D( \vec{J};s, \vec{b})$ is the unitary coherent-state
displacement operator defined as
\begin{equation}
D( \vec{J};s, \vec{b}) = exp[ \sum_{c }\int dq
\vec{J_{c}}(s, \vec{b};q) \vec{a_{c}}^{ \dagger}(q) -H.c.],
\end{equation}
where $ \vec{a_{c}}^{ \dagger}(q)$ is the creation operator of
a cluster c and the summation $ \sum_{c}$ is over all clusters.
The clusters are assumed to decay independently into
$ c = 1,2, \ldots $ pions
and outside the region of strong interactions.

Clusters decaying into two or more pions simulate a short-range correlation
between pions. They need not be
well-defined resonances. The more pions in a cluster, the
larger the correlation effect expected.

If the conservation of
isospin is a global property of the colliding system,
then $ \vec{J_{c}}(s, \vec{b};q)$ is of the form
\begin{equation}
\vec{J_{c}}(s, \vec{b};q) = J_{c}(s, \vec{b};q) \vec{e},
\end{equation}
where $ \vec{e}$ is a fixed unit vector in isospace independent of q.
The global conservation of isospin thus introduces the long-range correlation
between the emitted pions.
\newpage
\baselineskip=24pt
\section{Distribution of Pions in Isospace}
If the isospin  of two incoming
particles is $II_{3}$ , then the initial-state
vector of the pion field is $ \hat{S}(s, \vec{b}) \mid II_{3} \rangle,$
where $ \mid II_{3} \rangle$ is a vacuum state with no pions.
The $n$-pion production amplitude is
\begin{equation}
iT_{n}(s, \vec{b};q_{1} \ldots q_{n}) = 2s \langle I'I'_{3};q_{1} \ldots
q_{n} \mid \hat{S}(s, \vec{b}) \mid II_{3} \rangle,
\end{equation}
where $ I'I'_{3}$ denotes isostate of the outgoing leading particles.
The unnormalized probability distribution of producing $n_{+} \pi^{+}, \,
n_{ \_} \pi^{ \_},$ and $n_{0} \pi^{0}$ pions is defined as
\begin{eqnarray}
W(n_{+}n_{ \_}n_{0}, I'I'_{3},II_{3}) & = & \int d^{2}bdq_{1}dq_{2}
\ldots dq_{n} \mid \langle I'I_{3}n_{+}n_{ \_}n_{0} \mid \hat{S}(s,
\vec{b}) \mid II_{3} \rangle \mid^{2}, \\
 \mbox{where} \hspace{2.5cm} n & = & n_{+} + n_{ \_} + n_{0}. \nonumber
\end{eqnarray}

Assuming further that all $(I', I_{3}')$ are produced with
equal probability, we can sum over all possible isospin states of
the outgoing leading particles to obtain \\
\begin{equation}
P_{II_{3}}(n_{+}n \_ n_{0} ) = \frac{
\sum_{I'I'_{3}}W(n_{+}n\_ n_{0},I'I'_{3};II_{3})}{
\sum_{n_{+}n \_ n_{0}} \sum_{I'I'_{3}} W(n_{+} n \_
n_{0}, I'I'_{3};I I_{3} )}.
\end{equation} \\
as our basic relation for calculating various pion
distributions, pion multiplicities, and pion
correlations between definite charge combinations.

In order to obtain some more detailed results for multiplicity
distributions and correlations, one should have an explicit
form for the source function $J_{c}(s, \vec{b};q), c=1,2, \ldots \,.$

We shall analyze the isospin structure of our model
in the so called grey-disk model in which
\begin{equation}
\overline{n}_{c}(s, \vec{b}) = \overline{n}_{c}(s) \theta(b_{o}(s)-b),
\end{equation}
where $ \overline{n}_{c}(s)$ denotes the mean number of clusters
of the type $c$; $b_{0}(s)$ is related to the total inelastic
cross section, and $ \theta$ is a step function.
\newpage
\baselineskip=24pt
\section{Correlation between Neutral Pions}
We assume that pions are produced both directly and through isovector
clusters of the $ \rho$-type.

In order to calculate the first two moments of the joint probability
distribution $ P_{II_{3}}(n_{0},n_{ \_}) $ , we define the
generating function $ G_{II_{3}}(z,n_{ \_}) $
\begin{equation}
G_{II_{3}}(z,n_{ \_}) = \sum_{n_{0},n_{+}} P_{II_{3}}(n_{+},n_{ \_},
n_{0})z^{n_{0}},
\end{equation}
from which we calculate
\begin{equation}
\langle n_{0} \rangle _{n_{ \_}} = \frac{d}{dz}lnG_{II_{3}}(1,n_{ \_}),
\end{equation}
\begin{equation}
f_{2,n_{ \_}}^{0} = \frac{d^{2}}{dz^{2}}lnG_{II_{3}}(1,n_{ \_}).
\end{equation}
The form of the generating function $ G_{II_{3}}(z,n_{ \_}) $is
the following
\begin{equation}
G_{II_{3}}(z,n_{ \_})  =  (I + \frac{1}{2})
 \frac{(I-I_{3})!}{(I+I_{3})!}
\int_{-1}^{1}dx \mid P_{I}^{I_{3}}(x) \mid^{2} \frac{ \textstyle
A(z,x)^{n_{0}}}{ \textstyle n_{0}!}e^{ \textstyle - B(z,x)},
\end{equation}
where
\begin{equation}
2A(z,x) = (1-x^{2}) \overline{n}_{ \pi}+z(1-x^{2}) \overline{n}_{ \rho}+
2x^{2} \overline{n}_{ \rho}
\end{equation}
and
\begin{equation}
2B(z,x) = \overline{n}_{ \pi}(1+x^{2} - 2zx^{2}) +
 \overline{n}_{ \rho} (2 - z(1-x^{2})).
\end{equation}
Here $ \overline{n}_{ \pi}$ denotes the average number of directly
produced pions, and $ \overline{n}_{ \rho}$ denotes the average number
of $ \rho$-type clusters which decay into two short-range correlated
pions. The function $P_{I}^{I_{3}}(x)$ denotes the associate
Legendre polinomial. Note that $A(1,x)=B(1,x)$.

The total number of emitted pions is
\begin{equation}
\langle n \rangle = \overline{n}_{ \pi} + 2 \overline{n}_{ \rho}.
\end{equation}

In Fig. 1 we show the behaviour of $ \langle n_{0} \rangle _{n_{ \_}}$
for different combination of $( \overline{n}_{ \pi},
\overline{n}_{ \rho})$ when  $I=I_{3}=1.$
For Centauro-type behavior to appear the slope of
$ \langle n_{0} \rangle _{n_{ \_}}$ should be negative. However this
is only possible if
$ \overline{n}_{ \rho}=0.$ Recent estimate of the ratio of
$ \rho$-- mesons to pions at accelerator energies is
$ \overline{n}_{ \rho} = 0.10  \overline{n}_{ \pi}$.

In Fig. 2 we show the behaviour of the dispersion $D(n_{0})_{n_{ \_}}$
which is related to $f_{2,n_{ \_}}^{0}$ as
\begin{equation}
f_{2,n_{ \_}} = D(n_{0})_{n_{ \_}}^{2} -
\langle n_{0} \rangle _{n_{ \_}}
\end{equation}
again for different pairs of
$( \overline{n}_{ \pi},\overline{n}_{ \rho})$ and $I=I_{3}=1.$
It should be interesting to measure $f_{2,n_{ \_}}$ as it is
a sensitive quatity of the pairing properties of the pions.
\newpage
\baselineskip=24pt
\section{Conclusion}
The results of the present analysis have shown that the
emission of isovector clusters of pions in the framework of
an unitary eikonal model with global conservation of
isospin supresses the large isospin fluctuations of the Centauro-type
for pions. This might suggest that Centauro-type effect, if
it exists, will probably appear only in very limited
regions of phase space where isovector clusters should be
missing. How it is possible dynamically is not clear to the
authors  [13--16].

{ \large \bf Acknowledgment }

This work was supported by the Ministry of Science of
Croatia under Contract No. 1 - 03 - 212.

\newpage

{\bf Figure captions :}

Fig. 1. The average number of neutral pions  as a function of
the number of negative pions for
$ \; I = I_{3} = 1.$ and $ \overline{n}_{ \pi}+ \overline{n}_{ \rho}
= 18.$
The curves represent different combinations of
$( \overline{n}_{ \pi},  \overline{n}_{ \rho}),$
the average number of directly produced pions and the average
number of $ \rho$-type clusters, respectively.

Fig. 2. The correlation function for two neutral pions as a function
of the number of negative pions  for $I = I_{3} =
1.$ The curves represent different combinations of
$( \overline{n}_{ \pi}, \overline{n}_{ \rho}).$

\end{document}